# Keto-enol tautomerization drives the self-assembly of Leucoquinizarin on Au(111).

Roberto Costantini,[a,b] Luciano Colazzo,[c,d] Laura Batini,[e] Matus Stredansky,[a,b] Mohammed S. G. Mohammed,[c,d], Simona Achilli[e] Luca Floreano[b], Guido Fratesi[e], Dimas G. de Oteyza[c,d,f] and Albano Cossaro[b,*]

The self-assembly of Leucoquinizarin molecule on Au(111) surface is shown to be characterized by the molecules mostly in their keto-enolic tautomeric form, with evidences of their temporary switching to other tautomeric forms. This reveals a metastable chemistry of the assembled molecules, to be considered for their possible employment in the formation of more complex hetero-organic interfaces.

Hydroxyl or oxygen containing derivatives of anthracene are an important class of molecules, widely employed in industrial processes such as dye production[1,2], drugs synthesis[3], and others[4]. They are characterized by a fast redox response, which make them convenient building blocks for the design of cathodes of new generation of batteries[5–7] and for the synthesis of complex hybrid interfaces for photovoltaic devices[8]. Their color and in general their electronic properties can be finely tailored by varying the functionalization of the aromatic body[9]. The presence of one or more oxygen terminations promotes the occurrence of keto-enol tautomerism in these compounds, which can be of relevant importance for their functionality. For instance, in complexes formed by crown ethers and anthracene derivatives, the ion complexation of the former acts as a switch of the tautomerism of the anthracene group and promotes its luminescence, making the complexes suitable for the design of ion sensors [10–13]. As a matter of fact, most of the employed anthracene derivatives are based on 9,10-anthraquinone, an anthracene with double bonded oxygen atoms substituting the 9,10 positions. We report here a combined imaging, spectroscopic and theoretical study of the assembly on Au(111) surface of the Leucoquinizarin (LQZ, $C_{14}H_{10}O_4$), an anthracene with four hydroxyl groups in position 1,9,4,10 (fig 1), adopted in chemical industry for the synthesis of dyes. The molecule goes through keto-enol tautomerism and some of the tautomers are represented in Fig. 1. The presence of pairs of hydroxyl groups positioned in a double diol configuration makes LQZ a potential building block for extended covalent networks. In particular, one could exploit its tendency to aldol condensation with hydroxyl molecules,[14] as well as its covalent coupling with boronic molecules upon diol-boronic condensation reaction[15]. The latter perspective would be of interest due to the increasing importance the boronic chemistry is gaining in recent years as a tool for the synthesis of complex 2D interfaces[16–18]. In the view of exploring these synthesis routes to possibly exploit the electronic and optical properties of anthraquinone derivatives in more extended architectures, the chemistry of the LQZ on surfaces needs to be properly investigated first. Both experimental[19] and theoretical[20] findings indicate the T1 configuration as the most stable among the possible tautomers depicted in Fig. 1, both in gas phase and in solution. Our findings support T1 to be the LQZ form also on the Au(111) surface. The room temperature growth of LQZ sub-monolayer is characterized by differently ordered assemblies. The imaging of the most ordered phase is reported in Fig. 2 and exhibits a molecular assembly characterized by the presence of chiral pores, spaced by 2.5 nm. The morphology resembles the giant honeycomb architecture found for the anthraquinone molecule on Cu(111)[21]. Here however, the presence of additional oxygen terminations with respect to anthraquinone likely promotes a

a. Physics Department of University of Trietse, via A. Valerio 2, 34127 Trieste, Italy.
b. CNR-IOM, Area Science Park, Strada Statale 14, km 163,5, 34149 Trieste, Italy.
c. Donostia International Physics Center, Paseo Manuel Lardizabal 4, E-20018 Donostia – San Sebastián, Spain
d. Centro de Física de Materiales (CSIC-UPV/EHU) – MPC, Paseo Manuel de Lardizabal, 5 – E-20018 Donostia – San Sebastián, Spain
e. Dipartimento di Fisica "Aldo Pontremoli"Università degli Studi di Milano, Milano, Italy
f. Ikerbasque, Basque Foundation for Science, E-48011 Bilbao, Spain
* E-mail: cossaro@iom.cnr.it

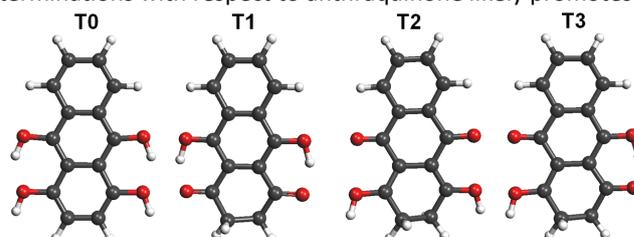

**Fig. 1**. Molecular structure of four tautomers of LQZ molecule.

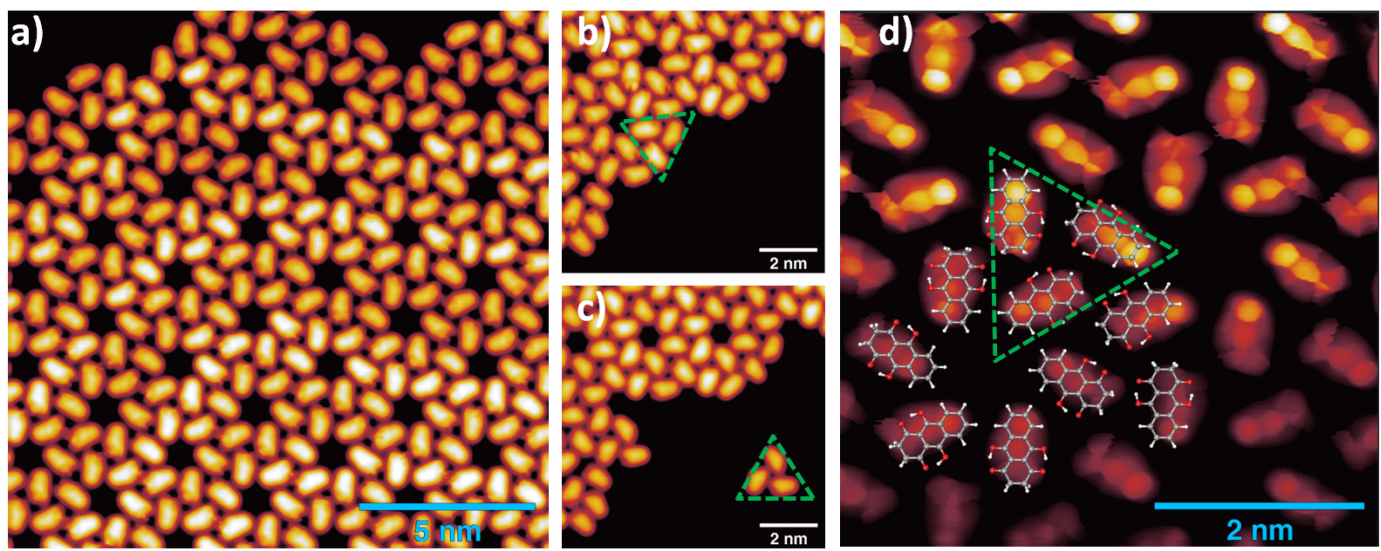

**Fig. 2** a) STM images of a LQZ submonolayer showing the presence of nanometric pores. (0.05V, 100pA) b) and c) Tip manipulation allows to separate a trimer from the assembly, which is identified as building block of the architecture. (0.1 V, 50 pA) d) bond resolution image reveals a distorted termination of the LQZ molecules, which can be related to the diketo phenol of T1 tautomer. The superimposed overlayer model results from the DFT calculations of the assembly, obtained as detailed in the ESI.

more compact assembly Domains with both chiralities are observed on the same surface. The supramolecular structure corresponds to a hierarchical assembly with molecular trimers as building units, which remain together upon controlled molecular manipulation with the scanning probe. (Fig. 2b and 2c). As already mentioned, the ordered assembly of Fig. 2, observed for low coverage (< 0.5 ML), is not the only morphology obtained in the preparations. At higher coverage, or with larger deposition rate, the nanopore domains become less regular due to an increase of the molecular density, finally yielding a disordered phase at completion of the first layer (ESI, Fig. S1). Interestingly, molecular trimers, even though characterized by a less regular geometry than those reported in Fig. 2, can still be identified as the recurring motif of the architectures. High resolution imaging with a CO functionalized tip is shown in Fig. 2d and allows observing more details of the LQZ assembly and of the intermolecular interactions. All the molecules display the same aspect: one bright hexagonal carbon ring at one end, a darker and more irregular hexagonal ring in the center, and an even darker and strongly deformed structure at the opposite end. We performed DFT calculations in order to have better insight into both the assembly and the electronic properties of the LQZ molecule. The details of the calculations and of the resulting structures are reported in the ESI. Our findings confirm the conformer T1 as the most stable form, with the purely hydroxylic T0 having 0.9 eV higher total energy in the gas phase. Moreover, the diketo form of the terminal quinol of T1 introduces a distortion of the anthracene backbone. This is likely the cause of the blurred aspect of one of the molecules terminations as imaged in Fig. 2d and suggests an assembly scheme with the aromatic ring (opposite to quinol ring) pointing towards the pores. The superimposed molecular model in Fig. 2d is the result of DFT relaxation of a free-standing LQZ layer, as obtained starting from the structure suggested by STM. The analysis of the intermolecular bonding confirms that the oxygen atoms have a determinant role in the assembly. In particular, the interaction between the keto-enolic terminations of molecules belonging to adjacent trimers drives the assembly of the porous architecture (see ESI for the details).

A further important aspect of LQZ assemblies, observed by STM measurements, is their unstable imaging. This is shown by way of example in Fig. 3. Panel b) displays the image of a disordered chain of molecules, measured on a low coverage LQZ layer deposited on Au(111) held at 120 K. Maintaining the probe at the point marked by the blue cross and monitoring the current under open feedback conditions provides a trace as shown in panel a), characterized by the presence of telegraph noise that is absent on the bare surface. By comparison with similar observations in previous studies, this effect can be related to the switching of the molecule to different tautomeric forms[22–24], although switching to different adsorption configurations cannot be excluded either. The curve exhibits two switching levels beside the most stable one, as evidenced by the blue dotted lines in the graph.

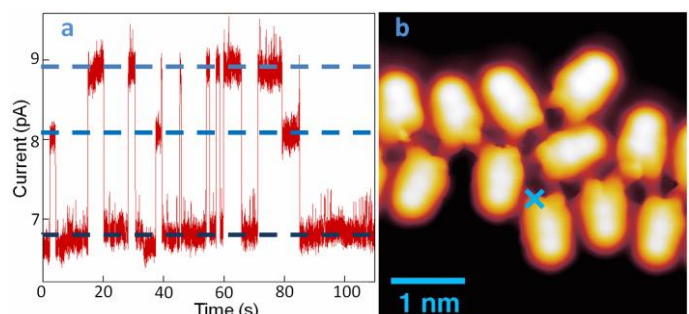

**Fig. 3** a) Tunneling current measured in correspondence of the cross indicated in the image (panel b, -0.15V, 1nA) of a very low coverage LQZ layer, as a function of the measuring time. Three different current levels are measured, suggesting a dynamics of the LQZ chemistry in the assembly, possibly due to tautomeric switching.

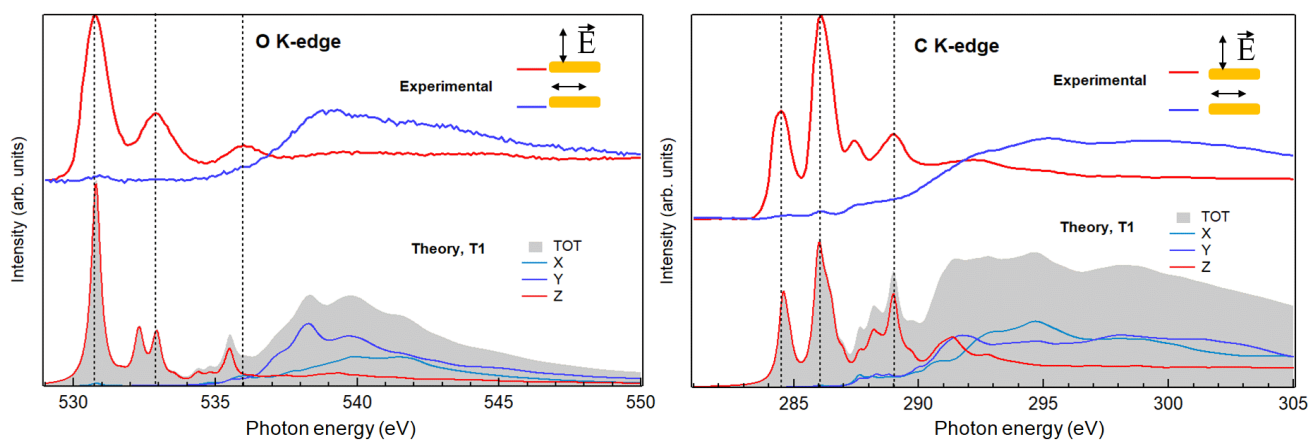

**Fig. 4**. O1s and C1s XPS spectra measured on the LQZ monolayer (top panels) and simulated for the T1 tautomer (bottom).

However, the switching rate and the different level's relative dominance greatly vary as the tip is positioned differently (Fig. S4). While this proves the molecule-related origin of the switching, it also reveals the influence of the location-dependent tip-molecule interactions, in line with other well-known tautomerization switches revealed in porphycenes.[24] Another similarity between the two systems is the promotion of the switching events by vibrational excitations of the molecule, which becomes obvious from the notably increased instabilities at bias values above ±50 meV, and a further increase at around ±350 meV. The symmetric thresholds for both bias polarities are a fingerprint of inelastic tunneling processes that we associate, in the absence of electronic orbitals and transitions at these low energies, to molecular vibrational modes. It should be noted, however, that instabilities in the tunneling gap are observed even at 1 meV (Fig. 3a), whereby tip-induced effects driven by inelastic tunneling excitations are minimized. The DFT energy difference between T1 and the other tautomers, as calculated for a free molecule, is too large for a tautomeric switch to be at the origin of the tunneling current behavior presented in Fig. 3, which has been measured at 4.3 K. However, it has to be considered that experimentally, both the intermolecular hydrogen bonding and the interaction with the substrate can considerably lower this energy barrier. Moreover, transitions to intermediate tautomers, where only one of the enolic terminations of T1 has switched, can be considered at the origin of the telegraph noise, with consequent reduced effort in terms of energy barrier.

To further investigate the chemistry and the morphology of the system we performed X-ray spectroscopy measurements at the ALOISA beamline[25] and at its ANCHOR-SUNDYN endstation[26] at the Elettra Synchrotron. Fig. 4 reports the O and C K-edge NEXAFS spectra taken on a LQZ monolayer. Carbon spectra resembles the anthracene ones[27], apart from the intensity ratio between the first two resonances, at around 285 eV. Both edges present a strong dichroism, with the lower energy transitions having maximum intensity when the polarization of the electric field is perpendicular to the surface.

We calculated the eigenfunctions for the ground state of the four tautomers, as well as the density of states projected onto states of π and σ symmetry. The four conformers exhibit the same symmetry of electronic states around HOMO. In particular, HOMO, LUMO and LUMO+1 are π states, whereas σ-symmetry can be found for the HOMO-1 (see ESI). The almost perfect dichroism indicates therefore that molecules are adsorbed flat on the surface and that neither the interaction with the substrate, nor the hydrogen bonding scheme the molecules are involved in, introduce relevant distortion of the electronic structure. The lower panels of the Fig. 4 report the calculated spectra for T1 tautomer, which provides a better match to the data than other configurations. Photoemission spectroscopy (XPS) also corroborates the presence of T1 species on surface. The O1s and C1s peaks are shown in Fig. 5, as measured on a LQZ monolayer, and as calculated for the T1 molecule. No significant differences have been found in the XPS profiles taken on films with lower coverage. The O1s peak presents two broad components, of the same intensity, found at 531.3 eV and 532.7 eV, which can be assigned to carbonyl (=O) and to hydroxyl (-OH) oxygen species respectively[28]. The profile of C1s is more complex and can be fitted by five components. The calculated spectra for T1 nicely reproduce the experimental profiles and confirm the O1s assignment. Regarding the C1s peak, the highest energy component of C1s at ~289 eV is not reproduced by calculations and can be assigned to energy loss due to inner excitations of the molecule. The C1s components 3 and 4 are due to the carbon atoms linked to the hydroxyl and carbonyl oxygen atoms respectively. We ascribe both the broadness of O1s components and the minor discrepancies between experimental and calculated C1s profiles to the intermolecular bonds that are not considered by the calculation, as well as to the morphologic disorder of the assembly and to the possible presence of different tautomers on the surface. In this regard, we remark that, the X-ray spectroscopy measurements have been taken at room temperature, whereas STM has been performed at 4K. We can speculate that these conditions promote a relevant portion of LQZ molecules in a tautomeric

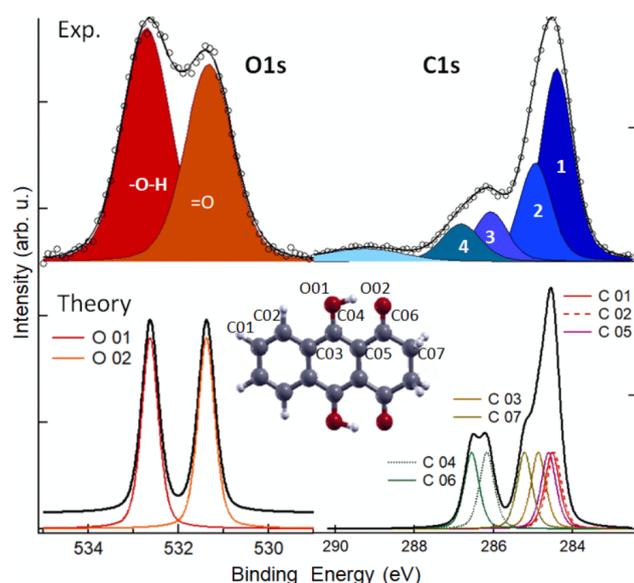

**Fig. 5**. XPS spectra measured (top) and calculated for T1 molecule (bottom). See the ESI for the comparison with the calculations of the other tautomers.

form different from T1.

In conclusion, the assembly of LQZ molecule on Au(111) has been characterized in terms of the morphology and of the electronic properties of the supramolecular architecture. A trimer of hydrogen bonded molecules has been identified as recurring motif of the assembly and as building block of differently ordered superstructures. The assembly scheme that can be inferred from the STM images on the highly ordered structures is consistent with the LQZ molecules predominantly in their keto-enolic tautomeric form. The same configuration is compatible with the X-ray spectroscopy results, as simulated by DFT calculations. We ascribe a certain degree of unpredictability of the morphology, as the molecular density increases, to the tautomeric switch the molecules can experience on the surface. This is corroborated by the observation of telegraph noise in the STS measurements. As different tautomers present different chemical properties of their oxygen terminations, our findings have to be taken into account in the perspective of employing LQZ as building block of more complex interfaces and of the possible on-surface reactions it can be involved in.

We acknowledge: the CINECA award under the ISCRA initiative, (Grant No. HP10CB0ZW2); the MIUR SIR grant SUNDYN (RBSI14G7TL, CUP B82I15000910001); funding from the European Union's Horizon 2020 programme (Grant Agreement No. 635919 ) and from the Spanish MINECO (Grant No. MAT2016-78293-C6).

## Conflicts of interest

There are no conflicts to declare.

## Notes and references

**Keto-enol tautomerization drives the self-assembly of Leucoquinizarin on Au(111).**

Roberto Costantini, Luciano Colazzo, Laura Batini, Matus Stredansky, Mohammed S. G. Mohammed, Simona Achilli, Luca Floreano, Guido Fratesi, Dimas G. de Oteyza and Albano Cossaro

# Supporting Information

## 1. Materials and methods

**Experimental.**

The LQZ powder was purchased from TCI (purity >98%) and sublimated from a Knudsen cell at 370K. Deposition was performed at room temperature. The Au(111) sample was cleaned by cycles of Ar+ sputtering and annealing at 800 K.

O1s and C1s XPS were measured at 650 eV and 515 eV with overall resolution of 0.25 eV and 0.2 eV respectively. Binding energy calibration was done by measuring the Au4$f_{7/2}$ peak and by aligning its bulk component to 84.0 eV[1]. NEXAFS spectra were acquired in Auger yield measuring the Auger signal at 252 eV and 507 eV respectively.

STM measurements have been performed at the *Donostia* International Physics Center on a commercial Scienta-Omicron LT-SPM system at 4.3 K. A mechanically clipped PtIr wire was used as tip, sharpened by voltage pulses and indentation into clean Au(111) patches. The images were analysed with the WSxM software.[2]

**Calculations.**
We performed the theoretical analysis based on density functional theory (DFT) simulations within the Perdew-Burke-Ernzerhof (PBE) approximation to the exchange-correlation functional[3], using the Quantum-ESPRESSO simulation package[4] and the same norm-conserving pseudopotentials and numerical setup as in our previous studies[5]. Free molecules have been optimized in periodically repeated orthorhombic cells with a vacuum separation between replicas of 11 Å. We evaluate core-level spectra as described elsewhere.[6] Self-consistent calculations with a C pseudopotential generated with a 1s full core hole (FCH)[7] at a given atom site provide the XPS core level shifts between inequivalent carbon (oxygen) atoms: those range from C01 to C07 (O01 to O02) for tautomers T0, T1, and T2, and up to C14 (O04) for T3. Next, we evaluate NEXAFS within the half-core-hole approach (HCH) [8,9] by using the xspectra code[10]. Within a pseudopotential approach, the XPS and NEXAFS spectra of the molecule are both defined up to an energy constant, which is adjusted to match the experimental features.

## 2. STM imaging

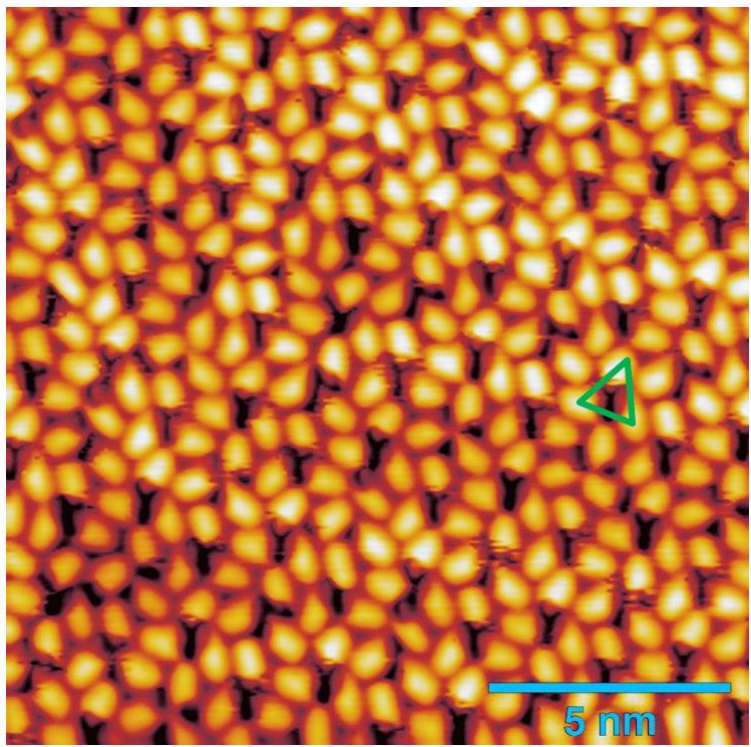

**Figure S1** The RT growth of the monolayer of LQZ leads to less ordered phases with respect to the honeycomb structure presented in the manuscript. As shown in the image, a quite compact assembly of molecules is formed, where molecular trimers similar to the ones discussed in the manuscript, exhibiting a more distorted geometry, can be identified (green triangle). STM parameters: 0.03 V, 200 pA.

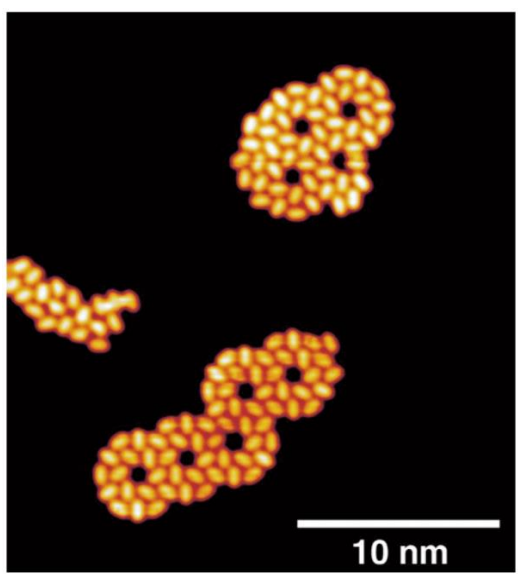

**Figure S2.** Thermal treatment of the LQZ films reveald that the desorption of the molecules is obtained at sample temperature Ts~500K. Image shows a very low coverage film, obtained upon annealing of a submonolayer at 420K, where both chiralities of the pores are present.( 0.05 V, 100 pA).

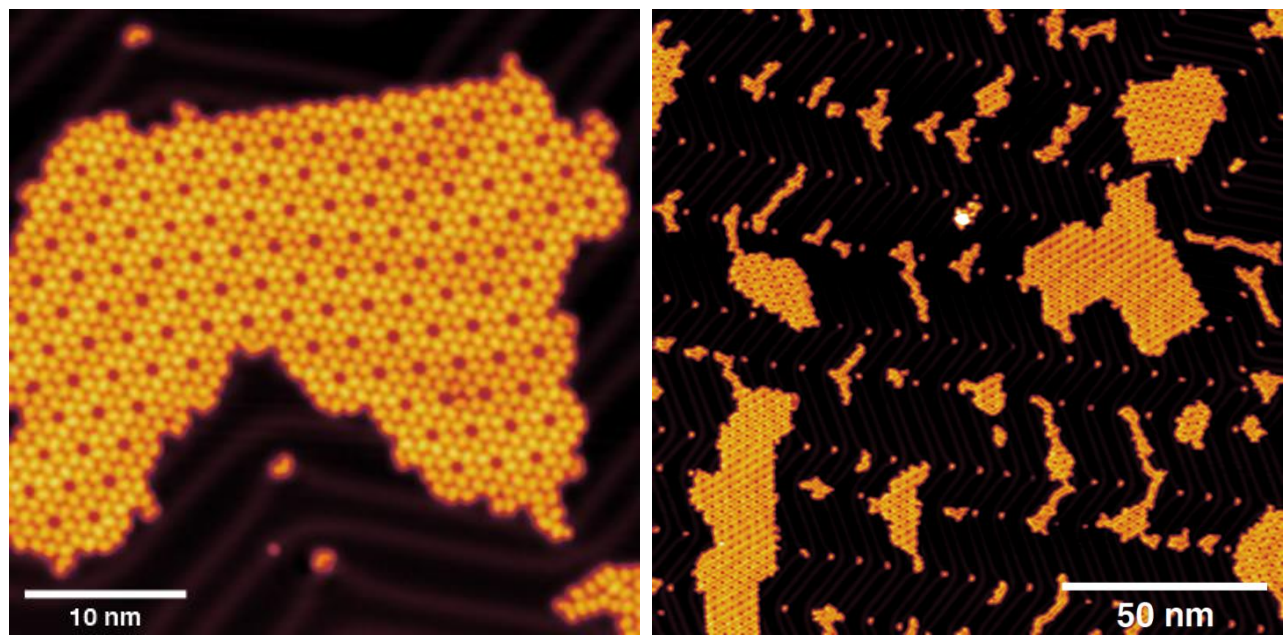

Fig S3 Two large scale images of the porous phase presented in the manuscript (0.4V, 20pA).

## 3. Telegraph noise

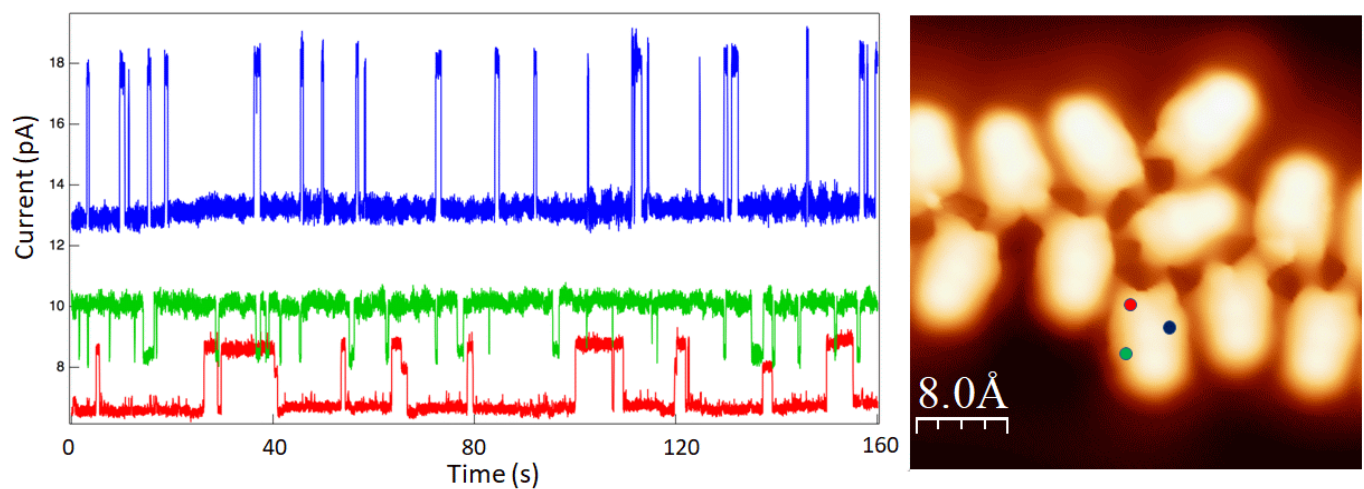

Figure S4. Three telegraph noise traces taken with different tip positioning, as indicated by the coloured circles in the image. The red curve is taken from the same measurement presented in figure 3 of the manuscript (last part of this curve corresponds to the first part of the curve of the manuscript, upon shifting the time scale). As stated in the manuscript, both rate and different level's dominance vary as the tip positioning is changed.

# 4. DFT calculations: properties of gas phase tautomers

The energies of tautomers in the gas phase and their frontier orbitals are reported in Table S1 and Figure S5, respectively. For simplicity, calculations of orbitals and spectra have been performed for molecules with reflection symmetry with respect to the *xy* plane. These tautomers are referred as TNsym. In the case of T1, the tautomer with lowest energy, it has been verified that this symmetry assumption does not significantly affect the results when calculating the orbitals allowing for the distortion of the anthracene backbone. In the following paragraphs it will be shown that the spectroscopy simulations are also weakly affected by this relaxation. The T2sym geometry we considered, with the hydroxyl hydrogen atoms pointing away from the central ketones, represents the stable configuration for this tautomer. In fact, the configuration in which the hydrogen atoms point towards the ketones spontaneously switched to T1sym while performing structural optimization.

| Tautomer | T0(sym) | T1sym | T1 | T2 sym | T2 | T3 sym | T3 |
|---|---|---|---|---|---|---|---|
| Energy relative to T1 (eV) | 0.90 | 0.09 | 0 | 2.10 | 1.98 | 1.09 | 1.01 |

**Table S1**. Calculated Total Energy for the four tautomers of Figure 1, in their symmetric form (T*N*sym) and in their relaxed form (T*N*). T0 preserves the symmetry after relaxation so T0sym=T0.

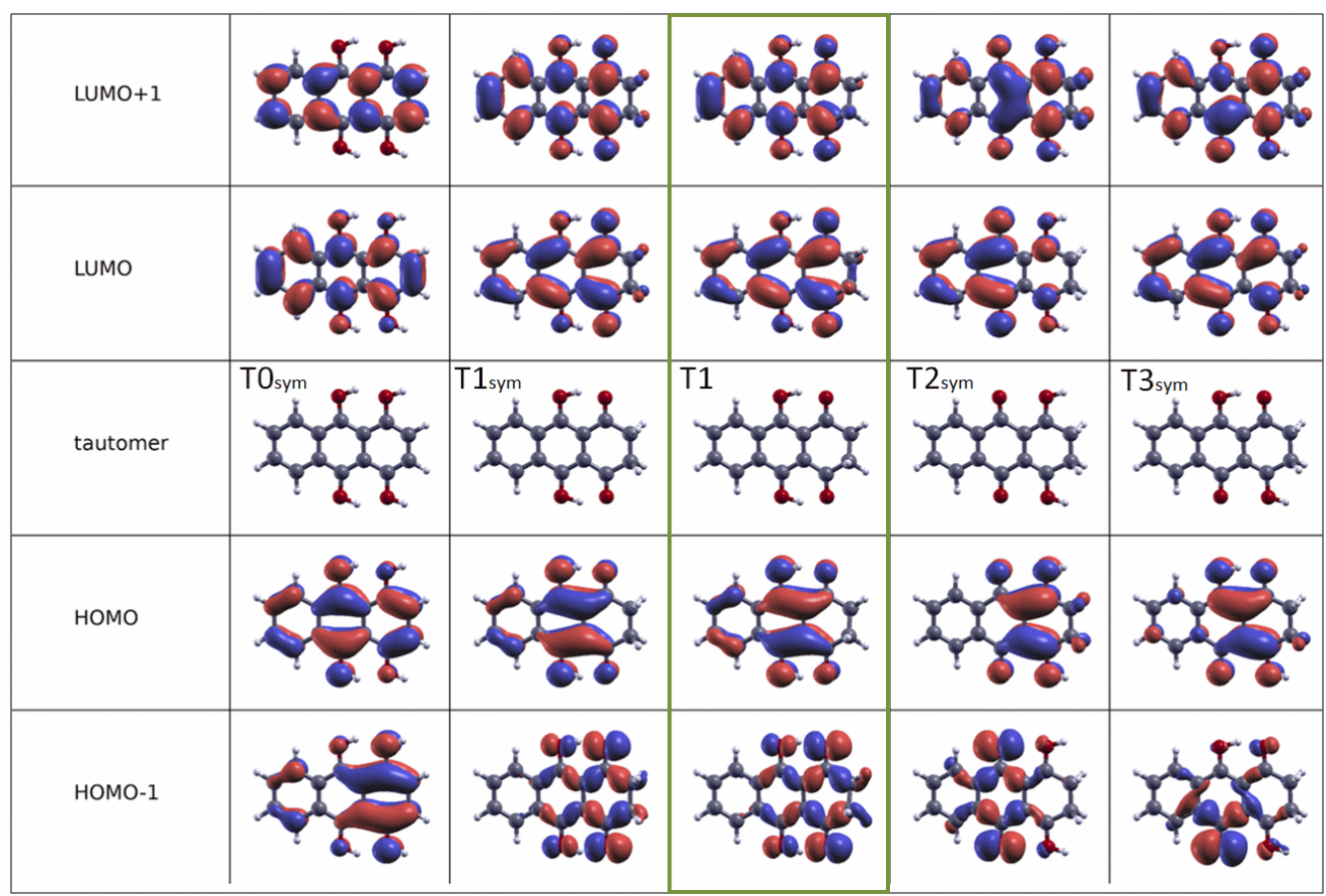

**Figure S5.** Graphical representation of the first occupied and unoccupied orbitals calculated for the different LQZ symmetric tautomers and for T1.

# 5. DFT calculations: Spectroscopic assignments

## 5.1: T1sym vs T1

The presence of the diketo-quinol upon tautomerization promotes the geometrical distortion of the corresponding termination of the anthracene backbone. Figure S6 reports the C1s NEXAFS and XPS calculations for the T1 tautomer in its distorted form (T1) and for the case in which the symmetry of the backbone was forced to be preserved (T1sym). Our findings show that the distortion introduced by the tautomerization poorly affects the spectroscopic properties. The same result was found for O1s (data not shown). For this reason, for simplicity, the calculations of the spectra for the comparison between the different tautomers have been done adopting the symmetric form of the molecules. We remark that the T1 spectra presented in the manuscript are calculated for the most stable distorted case.

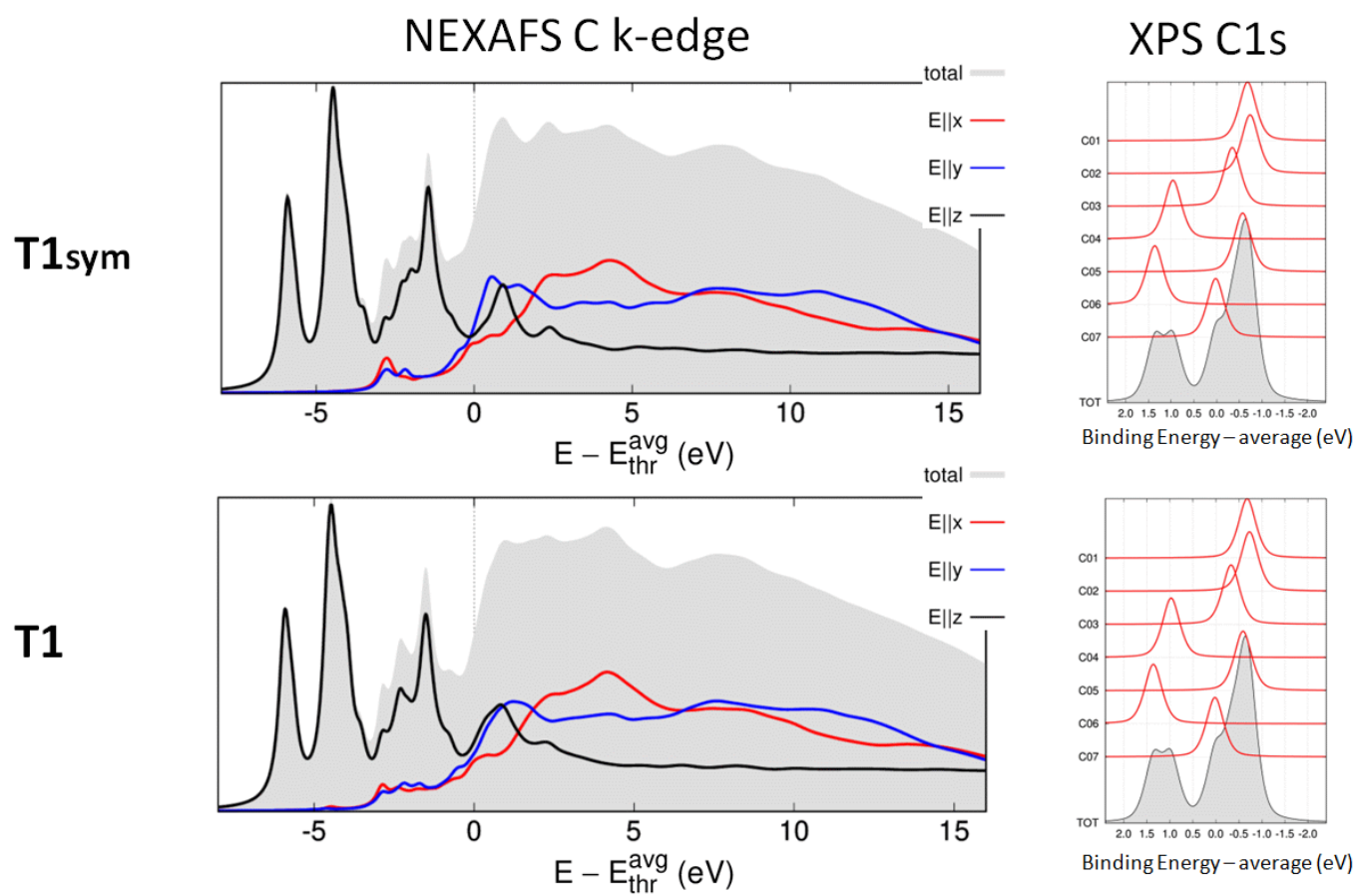

**Figure S6.** Calculated C1s spectra calculated for the distorted (top) and for the symmetric (bottom) T1 tautomer. The distortion only poorly affects the spectroscopic features on both NEXAFS and XPS spectra.

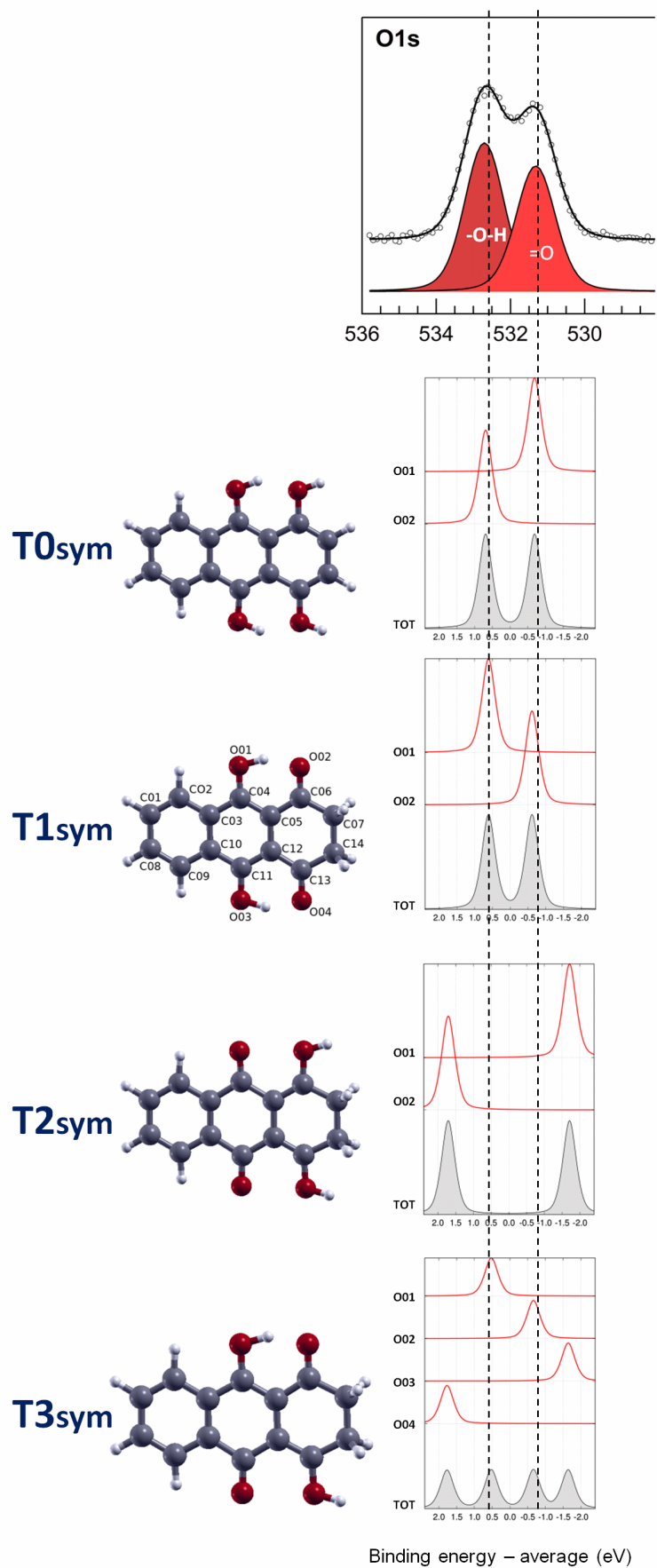

**Figure S7**. O1s XPS DFT simulated spectra for the T0sym, T1sym, T2sym and T3sym tautomers, compared with the experimental result. The simulated T1sym spectrum matches the experimental one, similarly to the T0sym spectrum. Conversely, features computed for T2sym and T3sym differ significantly from observations.

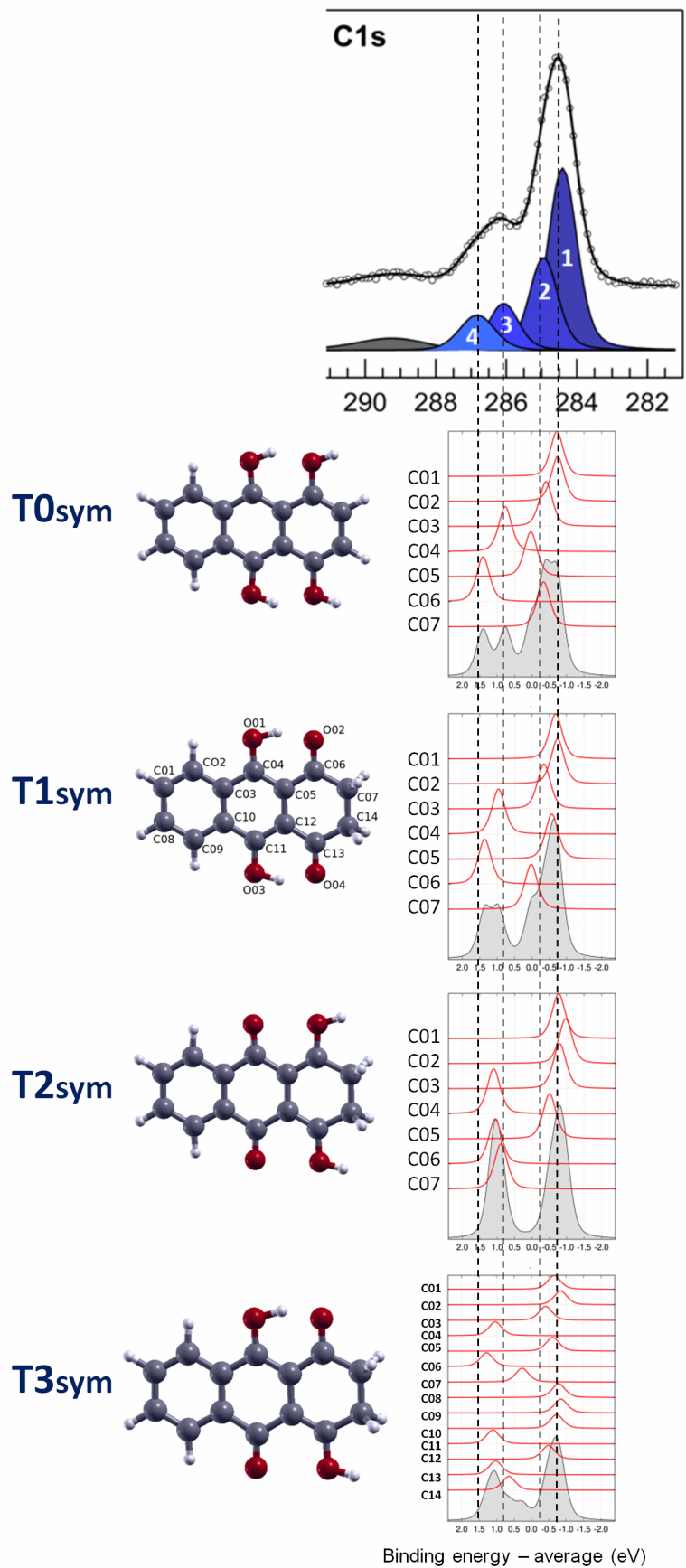

**Figure S8.** XPS DFT simulated spectra for the T0sym, T1sym, T2sym and T3sym tautomers, compared with the experimental result.

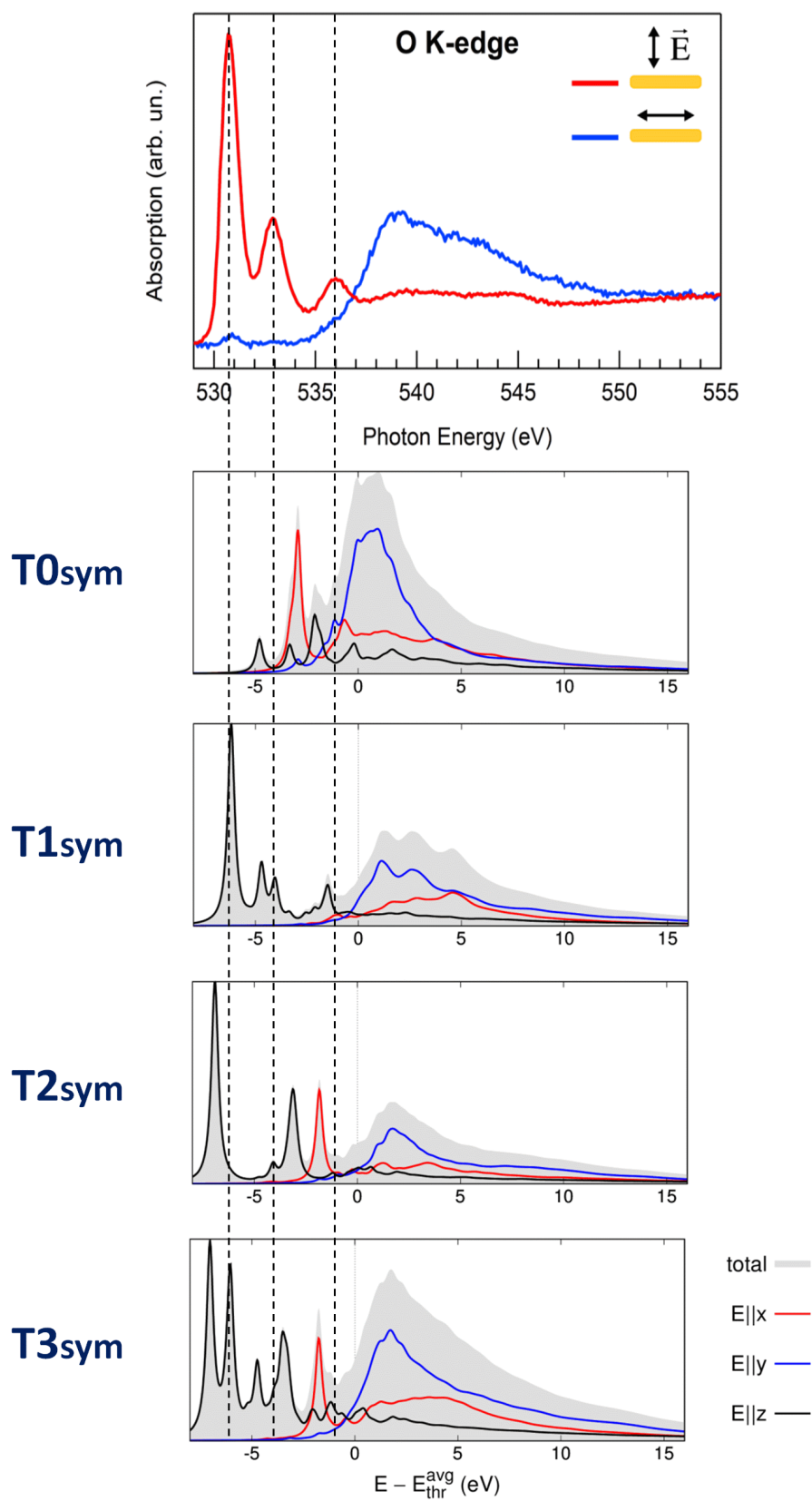

**Figure S9** O1s NEXAFS spectra measured (top panel) and simulated for the T0sym T1sym, T2sym and T3sym tautomers (lower panels). T1sym spectrum better agrees with experimental curve with respect to other tautomers, as it properly reproduces the three π* transitions indicated by the dotted black lines, in terms of both relative intensities and position. Energy alignment between theoretical and experimental curve is made by optimizing the overall superposition of the spectral features.

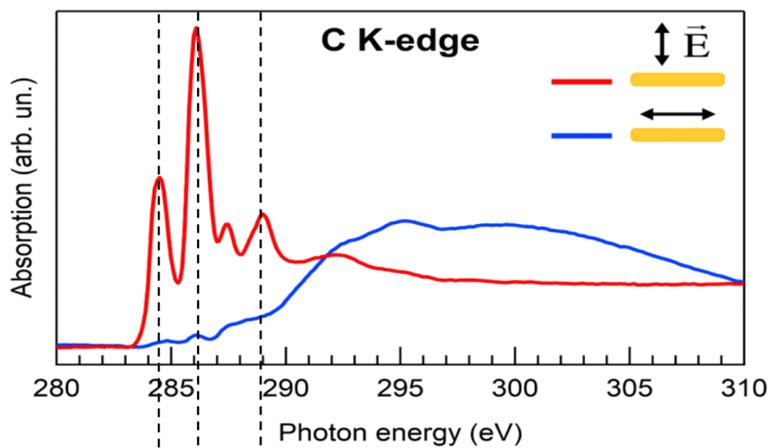
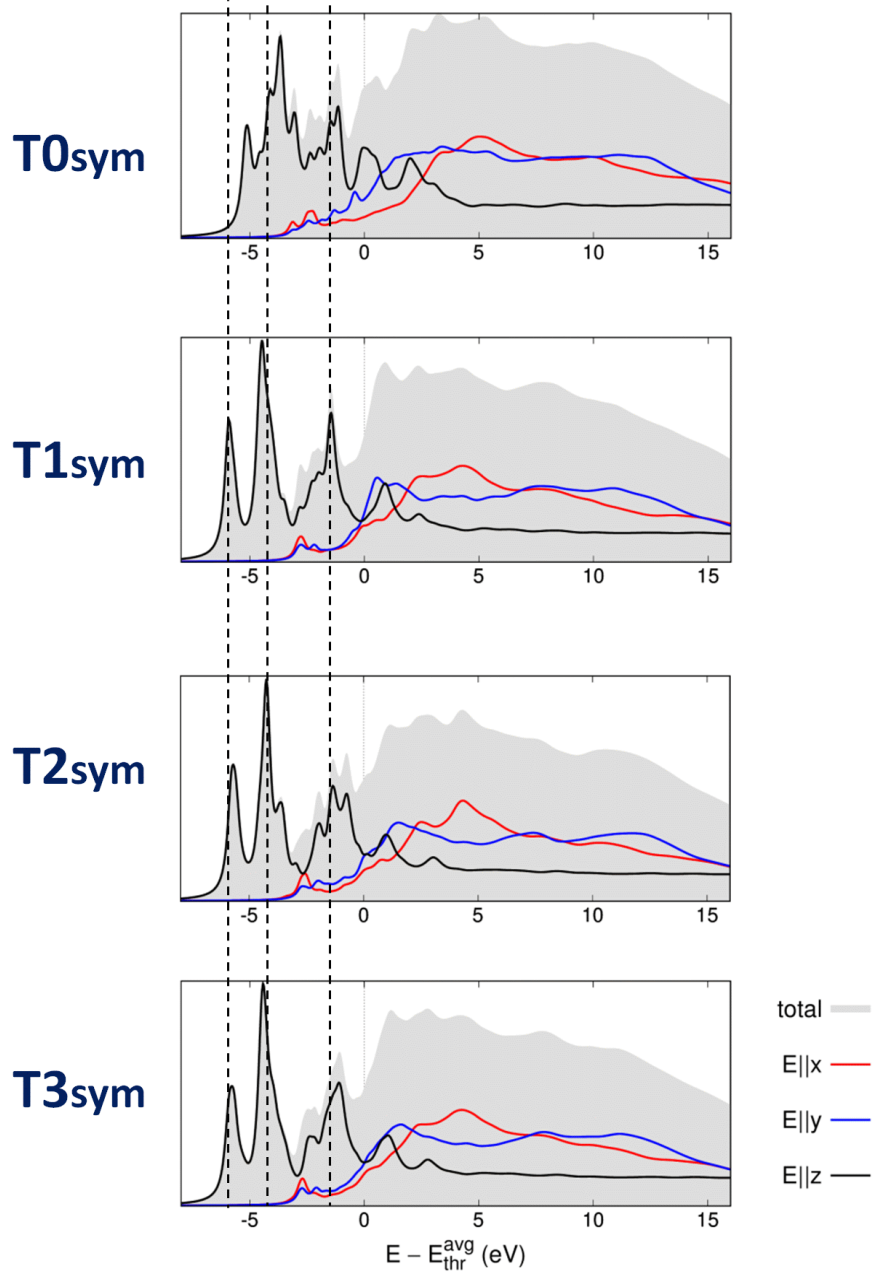

**Figure S10** C1s NEXAFS spectra measured (top panel) and simulated for the T0sym, T1sym, T2sym and T3sym tautomers (lower panels). As for O1s, T1sym spectrum better agrees with experimental curve with respect to other tautomers. Energy alignment between theoretical and experimental curve is made by optimizing the overall superposition of the spectral features.

# T1 tautomer: C1s Nexafs assignments

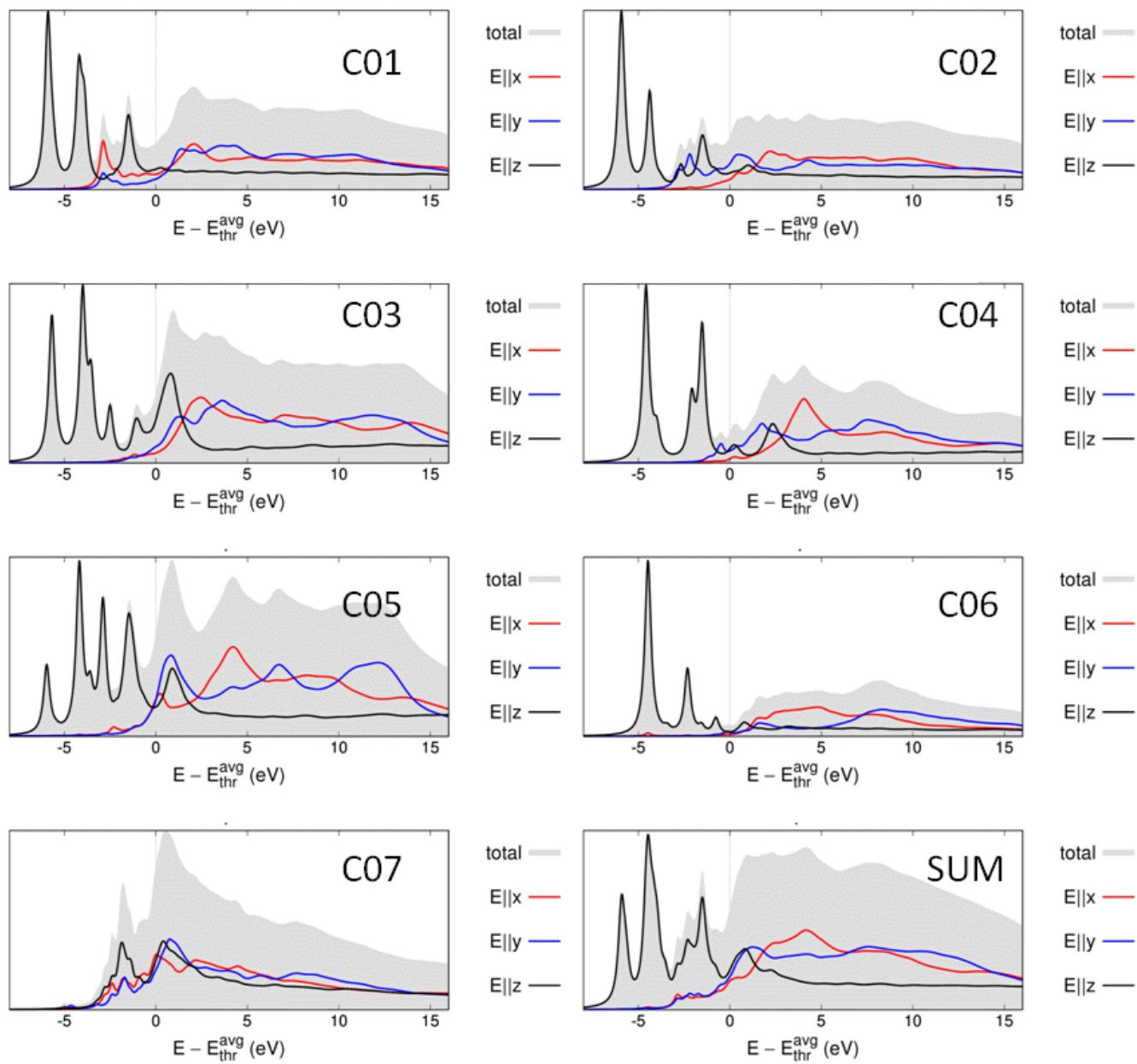

**Figure S11** Calculated T1 NEXAFS C1s spectral contributions of the non equivalent carbon atoms. See model in figure S6 for carbon atoms numbering.

## 6. DFT calculations: Intermolecular hydrogen bonding

In order to describe the intermolecular bonding scheme of the LQZ porous assembly, we built a model for a 2D crystal with 6 molecules / unit cell as inspired by the experimental STM images. The initial molecular coordinates are obtained from the gas phase structure by subsequent 60° rotations around a common axis located at the nanopore, about 10 Å from the center of the molecule. Variable-cell structural relaxations are then used to optimize the in-plane lattice constant and coordinates.

The unit cell of the overlayer contains two trimers (two of them indicated by the green dotted triangles in Figure S11). As a result of the calculations, the lattice constant of the molecular overlayer is 25.8 Å and the bonding energy is of 0.25 eV / cell. Each molecule (all are equivalent by symmetry) has 5 nearest neighbours, pairing in three different ways: (i) with two molecules within the same trimer; (ii) with one molecule lying nearly parallel; (iii) with two molecules facing the same nanopore. The resulting interaction scheme is depicted for one trimer in Fig S11. The coloured arrows identify, by proximity, the molecular terminations involved in the different interactions. In order to estimate the different interaction energies, we calculate the total energy of molecular overlayers where only pairs or trimers of the molecules of the unit cell involved in the considered interaction are included.

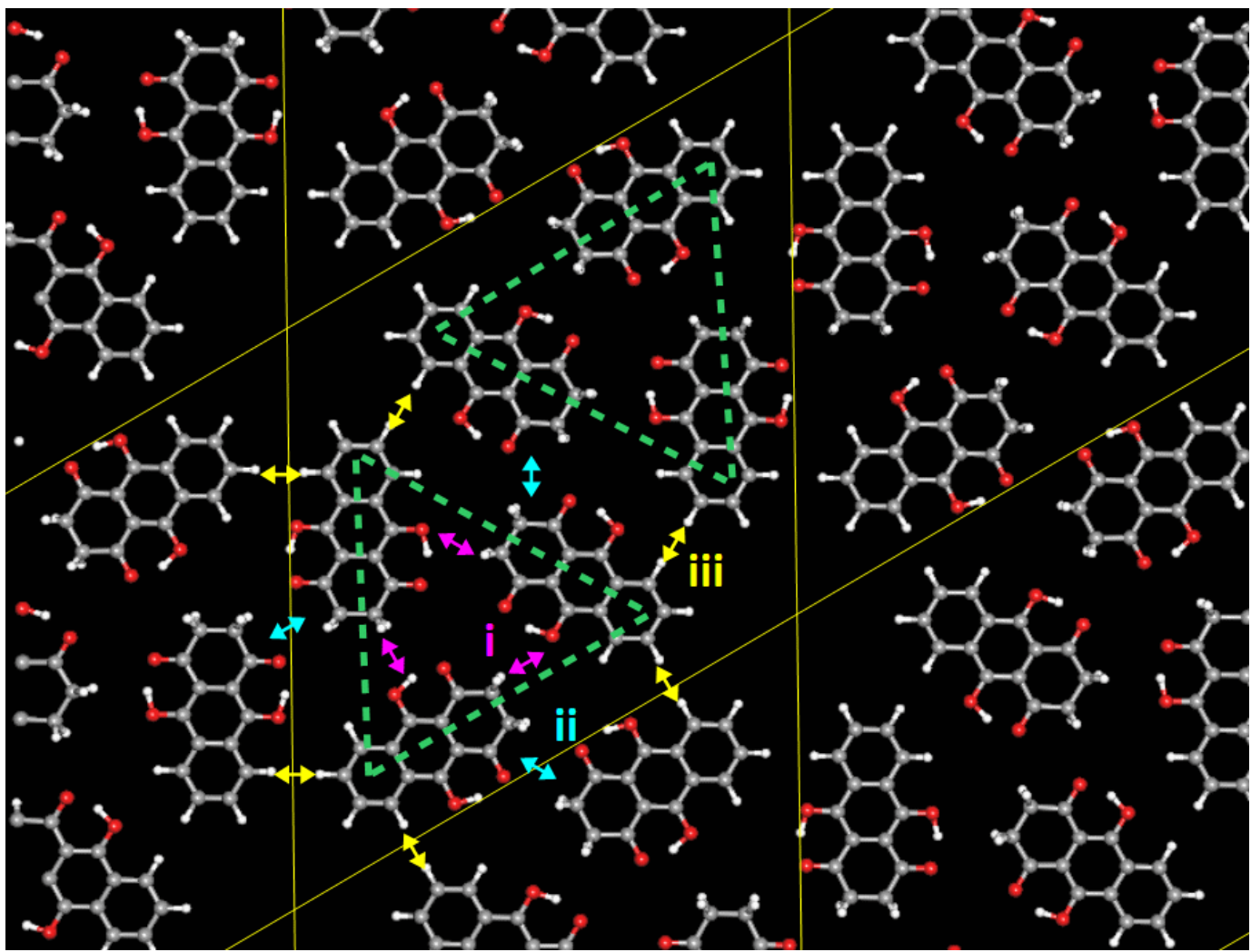

**Figure S11**. Structure of the molecular overlayer obtained by DFT calculations, by relaxation of the structure suggested by STM images

It results that : the trimers, which when isolated are stabilized by three (i) bonds, have a binding energy of 19 meV; the (iii) interaction in fact does not give any contribution to the bonding energy of the overlayer; the (ii) bonding between the trimers is 71 meV.

Both (i) and (ii) interactions are related to the oxygen atoms which, as expected play a fundamental role in the assembly process. The strongest intermolecular bond, (ii), is due to the facing of the keto-enolic termination of adjacent molecules. We can speculate that the formation of the trimers is promoted by the symmetry of the substrate and that the high (ii) affinity stabilizes clusters of trimers. On the other hand the (ii) interaction may be seen as a competitor of the trimer formation and lead to less ordered assembly of the LQZ molecules. The low coverage assembly reported in Figure S2 shows both situations, with most of the molecules forming regular chains of pores and part of them assembled in a disordered stripe. We recall however that we are neglecting the interaction of the molecules with the substrate, which could in part modify the relative importance of the different intermolecular interactions.